\documentclass[useAMS,usegraphicx,usenatbib]{mn2e} 
\bibpunct{(}{)}{;}{a}{}{,}
\def\fun{{\cal F}_\nu^{(1)}}
\def\fdeux{{\cal F}_\nu^{(2)}}

\def\kms{\,\mathrm{km\,s}^{-1}\,}
\def\kmskpc{\,\mathrm{km\,s}^{-1} \, \mathrm{kpc}^{-1}\,}
\def\tkmskpc{$\mathrm{km\,s}^{-1} \, \mathrm{kpc}^{-1}$}
\def\tkms{$\mathrm{km\,s}^{-1}$}
\def\kpc{\,\mathrm{kpc}\,}
\def\P{\mathbf{P}}
\def\mvr{$\,\langle V_R \rangle\,$} 
\def\mvrr0{\,\langle V_R \rangle_{R_0}\,} 

\title[The local spiral structure with RAVE]{The properties of the local
  spiral arms from RAVE data: two-dimensional density wave approach}

\author[A. Siebert et al.]{
A.~Siebert$^{1}$\thanks{E-mail: arnaud.siebert@astro.unistra.fr},
B.~Famaey$^{1}$,
J.~Binney$^{2}$,
B.~Burnett$^{2}$,
C.~Faure$^{1}$,
I.~Minchev$^{3}$,
\newauthor
M.E.K.~Williams$^{3}$,
O.~Bienaym\'e$^{1}$,
J.~Bland-Hawthorn$^{4}$,
C.~Boeche$^{5}$,
B.K.~Gibson$^{6,7}$,
\newauthor
E.K.~Grebel$^{5}$,
A.~Helmi$^{8}$,
A.~Just$^{5}$,
U.~Munari$^{9}$,
J.F.~Navarro$^{10}$,
Q.A.~Parker$^{11,12,13}$,
\newauthor
W.A.~Reid$^{11,12}$,
G. Seabroke$^{14}$,
A.~Siviero$^{15,3}$,
M.~Steinmetz$^{3}$
T.~Zwitter$^{16,17}$
\\
$^1$Observatoire  Astronomique, Universit\'e de Strasbourg,  CNRS, 11
rue de l'universit\'e, 67000, Strasbourg,  France \\
$^2$Rudolf Peierls Centre for Theoretical Physics, 1 Keble Road,
Oxford, OX1 3NP, UK\\
$^3$Leibniz-Institut f\"ur Astrophysik Potsdam (AIP), An der Sternwarte 16,
D-14482 Potsdam, Germany\\
$^4$Sydney Institute for Astronomy, University of Sydney, Sydney, NSW
2006, Australia. \\
$^{5}$Astronomisches Rechen-Institut, Zentrum f\"ur Astronomie der Universit\"at
Heidelberg, M\"onchhofstr. 12-14, D-69120 Heidelberg, Germany\\
$^6$Jeremiah Horrocks Institute, University of Central Lancashire,
Preston, PR1 2HE,UK\\
$^7$Department of Astronomy and Physics, Saint Mary's University,
Halifax, Nova Scotia, B3H 3C3, Canada\\
$^{8}$Kapteyn Astronomical Institute, University of Groningen, PO Box 800, 9700 AV Groningen, the Netherlands\\
$^{9}$INAF Astronomical Observatory of Padova, 36012 Asiago (VI), Italy\\
$^{10}$Department of Physics and Astronomy, University of Victoria,
Victoria, BC V8P 5C2, Canada\\
$^{11}$Department of Physics and Astronomy, Faculty of Science, Macquary
University, NSW 2109, Sydney, Australia\\
$^{12}$Macquarie Research Centre for Astronomy, Astrophysics and
Astrophotonics\\
$^{13}$Australian Astronomical Observatory, PO Box 296, Epping, NSW 2121, Australia\\
$^{14}$Mullard Space Science Laboratory, University College London, Holmbury St Mary, Dorking, RH5 6NT, UK\\
$^{15}$Department of Physics and Astronomy "G.Galilei", Padova University,
Vicolo dell'Osservatorio 2, Padova 35122, Italy \\
$^{16}$Faculty of Mathematics and Physics, University of Ljubljana,
Jadranska 19, SI-1000 Ljubljana, Slovenia\\
$^{17}$Center of Excellence SPACE-SI, Askerceva cesta 12, SI-1000
Ljubljana, Slovenia\\
 }

\begin{document}

\date{Accepted . Received ; in original form \today}

\maketitle

\begin{abstract}
  Using the RAVE survey, we recently brought to light a gradient in the mean
  galactocentric   radial  velocity   of   stars  in   the  extended   solar
  neighbourhood.   This  gradient  likely originates  from  non-axisymmetric
  perturbations of the potential, among  which a perturbation by spiral arms
  is a  possible explanation.  Here,  we apply the traditional  density wave
  theory and analytically model  the radial component of the two-dimensional
  velocity field.  Provided  that the radial velocity gradient  is caused by
  relatively  long-lived spiral  arms  that can  affect stars  substantially
  above the  plane, this analytic  model provides new  independent estimates
  for  the parameters  of  the  Milky Way  spiral  structure.  Our  analysis
  favours  a  two-armed  perturbation  with  the  Sun  close  to  the  inner
  ultra-harmonic     4:1     resonance,     with     a     pattern     speed
  $\Omega_p=18.6^{+0.3}_{-0.2}\kmskpc$     and     a     small     amplitude
  $A=0.55^{+0.02}_{-0.02}\%$  of  the  background  potential  (14\%  of  the
  background  density).  This  model  can  serve as  a  basis for  numerical
  simulations  in  three   dimensions,  additionally  including  a  possible
  influence of the galactic bar and/or other non-axisymmetric modes.
\end{abstract}


%
%
\section{Introduction}
\label{s:intro}

It has long been recognized that internal secular evolution processes should
play a major  role in shaping galaxy disks.  Among the  main drivers of this
secular evolution are the disc instabilities and associated non-axisymmetric
perturbations,  including the bar  and spiral  arms.  Questions  about their
nature          --           transient          or          quasi-stationary
\citep[e.g.,][]{sellwood2010,quillen2011,kawata}  -- , about  their detailed
structure and dynamics  such as their amplitude, pattern  speed, pitch angle
or number  of arms, as well  as questions about their  detailed influence on
secular  processes   like  stellar  migration   \citep{SB02,MF10},  are  all
essential  elements for a  better understanding  of galactic  evolution. The
Milky Way provides a unique laboratory  in which a snapshot of the dynamical
effect  of  present-day  disc  non-axisymmetries  can be  studied  in  great
detail, and help answering the above questions.

Current knowledge  of the structure and  dynamics of the bar  and the spiral
arms of the  Milky Way relies both  on the gas, and notably  on its observed
longitude-velocity   diagram   \citep{binney1991,bissantz2003,englmaier}  or
masers   in  high   mass  star   forming  regions   \citep[][and  references
therein]{reid2009},           and           on           the           stars
\citep[e.g.,][]{georgelin1976,binney1997,stanek1997,benjamin2005,lepine2011b}.
For the  spiral arms,  both types  of constraints are  combined in  a recent
study by \citet{vallee2008}, whose model  predicts the location in space and
velocity for the spiral arms.

With   the   advent   of   new  spectroscopic   and   astrometric   surveys,
six-dimensional phase-space  information for stars in  an increasingly large
volume  around the  Sun allow  us to  set new  dynamical constraints  on the
non-axisymmetric  perturbations of  the Galactic  potential.  An  example of
such new detailed kinematical information on stellar motions in the extended
solar  neighbourhood  is   the  recently  detected  (galactocentric)  radial
velocity  gradient of $\sim  4\kmskpc$ by  \citet{siebert11}, making  use of
more than 200  thousand stars from the RAVE survey.   If this result is
not owing  to systematic distance errors  (which the geometry  of the radial
velocity flow seems to exclude by not depending on distance and longitude in
any systematically biased  way), and more importantly, if  one assumes that,
at first order, what  is seen above the plane is a  reflection of what would
happen in  a razor-thin disc, and  that the spiral arms  are long-lived, one
can apply the  analytic density wave description of  spiral arms proposed by
\citet{linshu}  to constrain  the shape,  amplitude and  dynamics  of spiral
arms.

Whether  long-lived density  waves  are the  correct  description of  spiral
patterns in galaxies  remains heavily debated.  From a  theoretical point of
view, while it  seems that the radial velocity  dispersion profile needed to
support long-lived  spiral waves in  barless discs \citep[e.g.,][]{bertin96}
would be  heavily unstable \citep{sellwood2010}, the situation  is much less
clear  in the  presence  of a  central  bar, where  nonlinear mode  coupling
between  the  bar and  spiral  could  sustain  a long-lived  spiral  pattern
\citep{voglis2006, salo2010, quillen2011, minchev2012}, while \cite{kawata2}
however  find that  spiral arms  are  transient even  in the  presence of  a
central  bar. On  the  other hand,  \cite{Donghia}  find locally  long-lived
self-perpetuating spiral arms which could be locally consistent with density
waves,  but fluctuating  in  amplitude with  time.  Furthermore,  long-lived
spirals can also develop as  being sustained by coherent oscillations due to
a flyby galaxy encounter \citep{struck2011}, a process we know to be ongoing
for  the Milky  Way, and  cosmologically simulated  disk galaxies  exhibit a
distribution  of young stars  consistent with  the predictions  of classical
density wave theory for long-lived spirals \citep{pilkington2012}.  Finally,
let us  note that  both long-lived  and transient spirals  can coexist  in a
galaxy, which adds complexity to the picture.

On the observational  side the situation is also  unclear. Evidence seems to
exist for  both transient and  long-lived spiral arms: e.g.,  M81 apparently
contains long-lived  spiral arms consistent with the  classical density wave
theory \citep{lowe1994,  adler1996, kendall2008} while  in M51, even  if its
disc streaming motion appears  consistent with the density wave description,
the  mass fluxes  are inconsistent  with a  steady  flow \citep{shetty2007}.
Studying observational  tracers for  different stages of  the star-formation
sequence in 12 nearby spiral galaxies, \cite{foyle2011} also found that they
do not show  the expected spatial ordering for  long-lived spiral arms, from
upstream to  downstream in the corotating  frame. In the Milky  Way, many of
the  dynamical  constraints  on   spiral  arms  currently  come  from  local
constraints provided  by velocity space  substructures also known  as moving
groups \citep{dehnen1998,nimp}.   For instance, examining  the local stellar
distribution in  action space, \citet{sellwoodhyades} found  that stars from
the Hyades moving  group were concentrated along a  resonance line in action
space,  which was  interpreted as  a signature  of scattering  at  the inner
Lindblad resonance of a  transient spiral pattern \citep[see also][who found
that  this  feature  could  also   be  associated  with  an  outer  Lindblad
resonance]{mcmillanhyades}.   However,  in  this  picture, only  the  Hyades
moving group is  accounted for, and the remaining  substructures observed in
the  local phase  space distribution  must  be explained  by invoking  other
origins.      Models    based    only     on    transient     spiral    arms
\citep[e.g.,][]{desimone2004} were actually  unable to reproduce the precise
location of the  various other prominent moving groups,  such as Sirius.  On
the other  hand, models  based on long-lived  spiral arms, locating  the 4:1
inner resonance close  to the Sun, were able to  reproduce both the position
of   the   Hyades   and   Sirius    moving   groups   at   the   same   time
\citep{quillen2005,pompeia2011}   as    well   as   other    moving   groups
\citep{antoja2011}.   Other observational arguments  based on  the step-like
metallicity gradient in the Galactic  disc also argue in favor of long-lived
spirals \citep{lepine2011}.

Given  this  theoretical  and  observational  situation, we  here  make  the
conservative assumption  that interesting information can  be retrieved from
the classical analytic treatment of spiral arms as long-lived density waves.
This analytic model could then serve as a basis for numerical simulations in
a three-dimensional disc.  The paper is structured as  follows. In Section 2,
we review the data, as well as the analytic density wave model we are using.
We present and discuss our results in Section 3, and conclude in Section 4.

%
%
\section{Data \& Method}

\subsection{Two-dimensional velocity field}
\label{s:data}

Our analysis is based on data from the RAVE survey \citep{dr1,dr2,dr3} which
provides line-of-sight  velocities with a  precision of 2~\tkms for  a large
number of bright stars  in the southern hemisphere with $9 <  I < 12$.  RAVE
selects its targets  randomly in the I-band interval,  and so its properties
are  similar  to  a  magnitude   limited  survey.   The  RAVE  catalogue  is
cross-matched  with  astrometric  (PPMX,  UCAC2,  Tycho-2)  and  photometric
catalogues  (2MASS,   DENIS)  to  provide  additional   proper  motions  and
magnitudes.  In  this study,  we use the  internal version of  the catalogue
which contains  data for  434,807 spectra (393,903  stars).  To  compute the
galactocentric  velocities,  knowledge  of  the  distance  to  the  star  is
required.  For RAVE stars, distances to 30\% are available in three studies:
\citet{breddels2010},  \citet{zwitter2010}   and  \citet{burnett2011}.   All
catalogues  provide compatible  distance  estimators and  the velocity  maps
generated using the different catalogues are similar.

\begin{figure*}
\includegraphics[width=8cm]{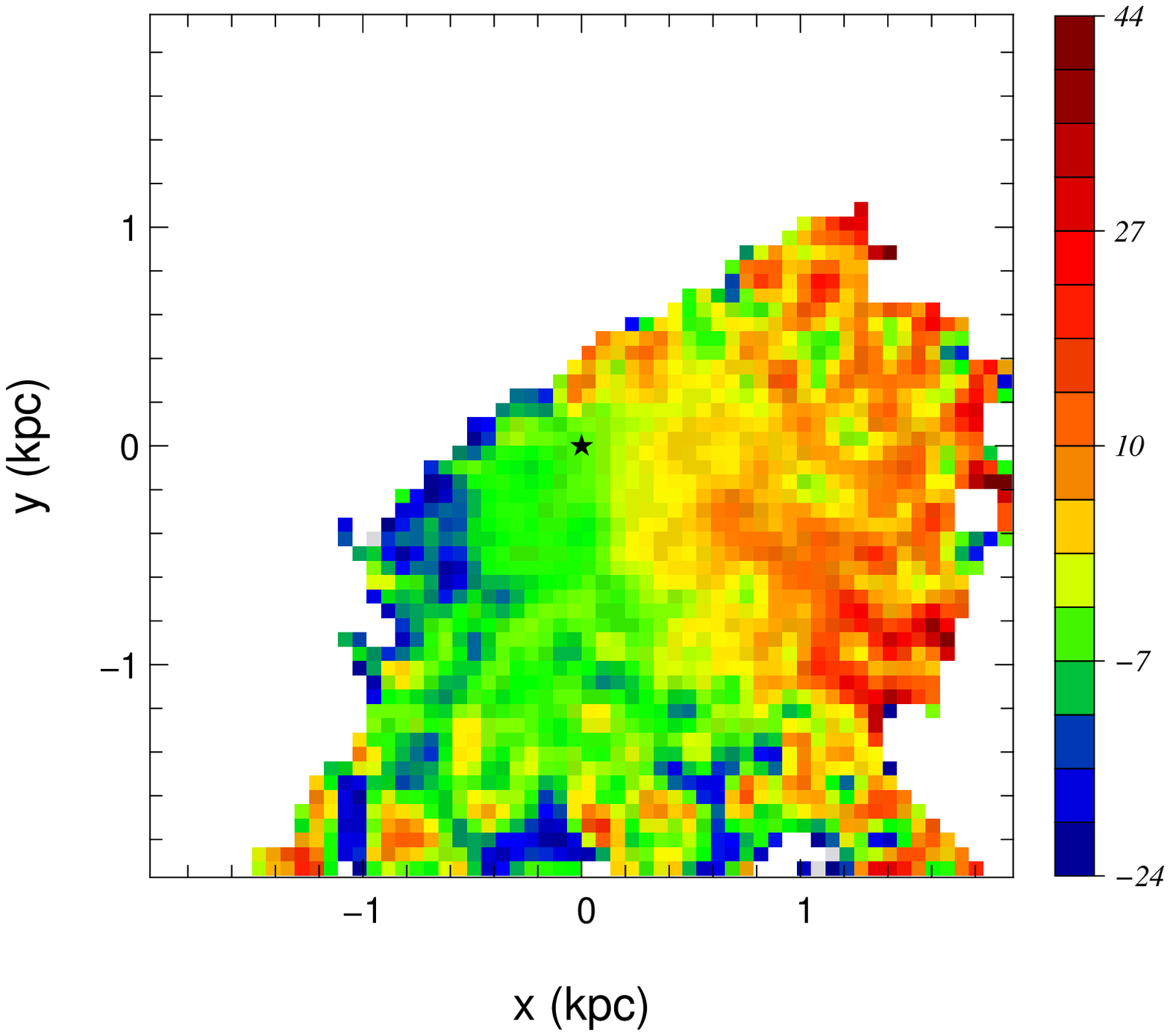}
\includegraphics[width=8cm]{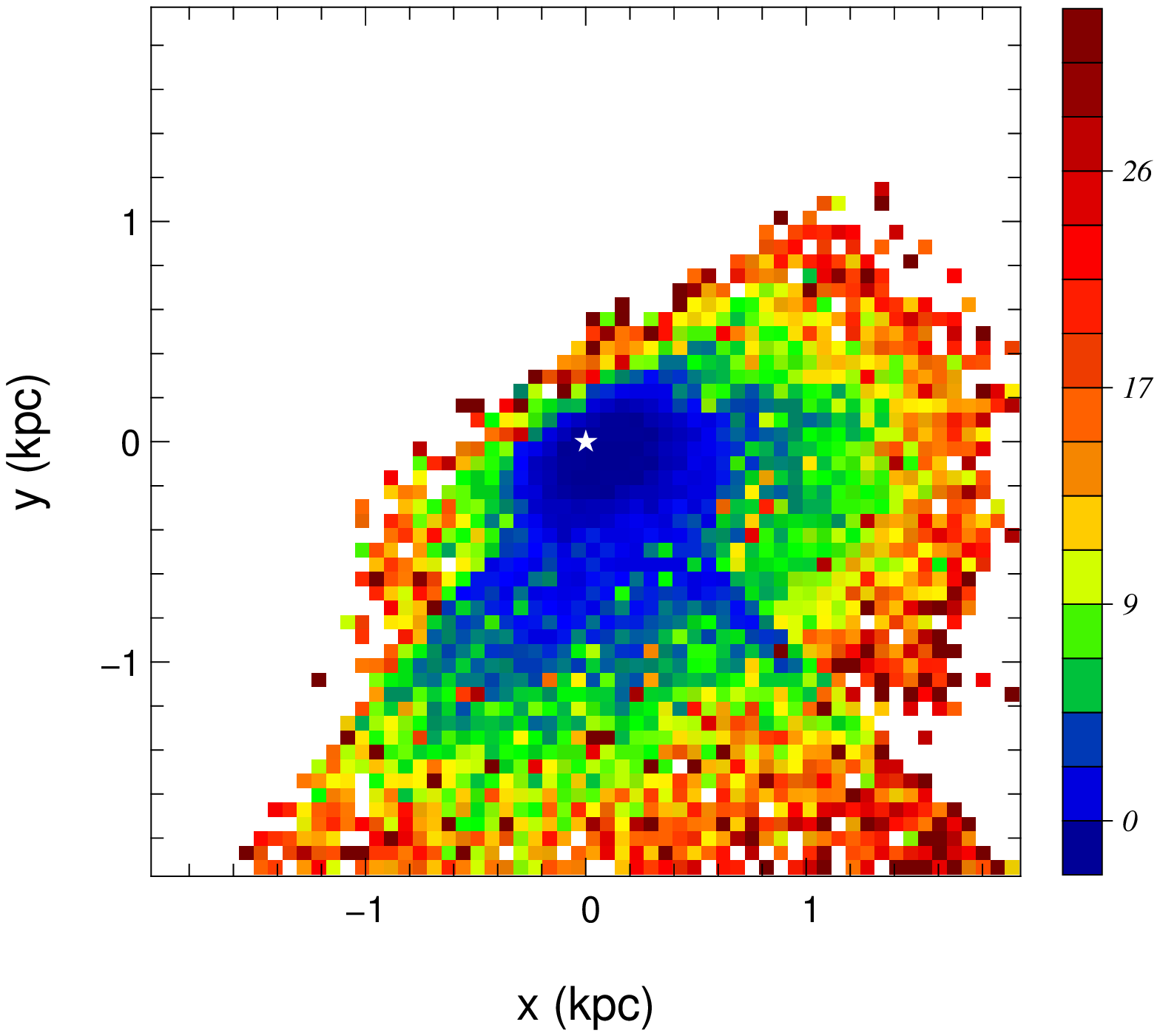}
\caption{Left:  velocity field  for  the radial  component  of the  velocity
  vector  \mvr as  a function  of location  in the  Galactic  plane.  Right:
  associated random error on the mean velocity. In both panels, the location
  of the Sun is marked by a  star, the Galactic centre is towards positive x
  and  the y-axis  is oriented  towards the  Galactic rotation.   The colour
  coding follows the mean velocity and mean velocity error in \tkms.}
\label{f:fields}
\end{figure*}

Our final  sample consists of  213,713 stars from  this survey limited  to a
distance of 2~kpc from the Sun and  to a height of 1~kpc above and below the
plane. We demonstrated the existence of a velocity gradient of disc stars in
the   fourth  quadrant,   directed   radially  from   the  Galactic   centre
\citep{siebert11}.  The two-dimensional  mean galactocentric radial velocity
field in the Galactic plane is presented in Fig.~\ref{f:fields} where we use
a box $4\times4$~kpc in size, centered  on the Sun, sampled using 60 bins in
each  directions.   For  this   analysis,  we  restrict  ourselves  to  bins
containing  more than  5 stars  and the  mean velocity  is computed  using a
median  function.   Note  that  converting velocities  in  the  heliocentric
reference  frame   into  the  $(V_R,V_\theta)$   galactocentric  coordinates
requires  the galactocentric  radius of  the Sun  $R_0$, the  Sun's peculiar
velocity  $\textbf{v}_\odot$ with  respect  to the  Local  Standard of  Rest
(LSR),  and the  motion  of the  LSR  with respect  to  the Galactic  center
$\textbf{v}_{\mathrm{LSR}}$.  We  assume $R_0=8  \,$kpc for the  distance of
the Sun to  the Galactic centre and $V_{\mathrm{LSR}}=220\kms$  to match the
values used for the mass  model (see Section~\ref{s:lin}). We use the latest
determination of  the value of  the solar motion  by \citet{schoenrich2010}:
$(U_\odot,V_\odot)=(11.1,12.24)\kms$.

The gradient  affects a sample dominated  at large distances  by red giants,
with a  typical velocity dispersion \mbox{$\sigma_R  \sim$ 30--40\tkms}, and
affects stars substantially above the plane, keeping in mind that RAVE lines
of sight are typically at $|b| >\sim 20^\circ$.  The zone where the gradient
is the steepest is populated with  stars with typically $|z| \sim 500 \,$pc.
However, if $U_{\mathrm{LSR}}$ is positive  (a local mean motion towards the
inner  Galaxy), then \mvr  would be  by construction  negative in  the Sun's
neighbourhood and reach  $0$ at larger distances and  larger heights. In the
modelling procedure hereafter, we let $\mvrr0 \equiv -U_{\mathrm{LSR}}$ be a
parameter of  the model  to allow  us to account  for uncertainties  on this
quantity.

Ideally one would  also use the tangential velocity  field $\langle V_\theta
\rangle$ in  combination to the \mvr  field.  However, as  stated above, our
sample  reaches distances  to  the  Galactic plane  of  1~kpc, avoiding  the
regions close  to the plane. The  RAVE survey mimicking  a magnitude limited
survey in fields  6$\,\deg$ in diameter on the sky,  each field containing a
different number of stars, the  stellar population mixture varies from point
to point on  the maps.  Therefore, the contribution  of the asymmetric drift
is difficult to  estimate while it enters the  calculation of the $V_\theta$
component.   Hence, we chose  to restrict  our analysis  to the  \mvr field,
although we give the full set of equations, including this component, in the
next section.

\subsection{Density Wave model}
\label{s:lin}

To model the  velocity field, we use the density  wave description of spiral
arms proposed  by \citet{linshu} and further developed  in \citet{lin69} and
\citet{shu71}.  This  model is based on  an asymptotic analysis  of the WKBJ
type of  the Euler/Boltzmann  equations, valid only  in the regime  of weak,
long-lived and tightly wound spirals  (small pitch angle).  This model being
well known  and documented \citep[see for example][]{BT08},  we restrict its
description to the main results used in this study.

The perturbation to the potential considered is of the form
\begin{equation}
\Phi_1=A(R)\exp \big[ i (\omega t - m \theta + \Phi(R))\big]\,,
\label{e:perturbation}
\end{equation}

\noindent  where $A(R)$ is  the amplitude  of the  perturbation, $m$  is the
number  of arms  and $\Phi(R)$  is a  monotonic function.   The perturbation
rotates at an angular frequency given by

\begin{equation}
\Omega_{\mathrm p}=\omega/m\,.
\label{e:pattern}
\end{equation}

The perturbations in the components of the mean velocities \mvr and $\langle
V_\theta \rangle$, where $(R,\theta)$ are the coordinates in the cylindrical
coordinate system centered on the Galaxy, are given by

\begin{eqnarray}
\langle V_R \rangle&=&\frac{k A}{\kappa} \frac{\nu}{1-\nu^2} \fun(x)
\cos(\chi) \nonumber\\
\langle V_\theta \rangle&=&-\frac{1}{2} \frac{k A}{\Omega} \frac{1}{1-\nu^2}
\fdeux(x) \sin( \chi )
\label{e:basis}
\end{eqnarray}

\noindent  where 
\begin{equation}
x=\frac{k^2 \sigma^2_R}{\kappa^2}\,,
\end{equation}

\noindent  $k$  being  the  radial  wave  number,  $\sigma_R$  the  velocity
dispersion, $\kappa$ the epicyclic frequency  and $\nu$ is defined by $\nu =
m(\Omega_{\mathrm p}-\Omega)/\kappa$.

The functions $\fun$ and $\fdeux$ are the ``reduction factors'' that correct
the  mean velocities  for the  effect of  velocity dispersion,  lowering the
effect  of the spiral  perturbation on  the velocity  field as  the velocity
dispersion  increases.   In  the  limit   of  a  zero   velocity  dispersion
($\fun=\fdeux=1$)  we  recover  the  velocity  field of  the  gas  while  if
$\sigma_R$ becomes large, $\fun=\fdeux\rightarrow  0$ and the velocity field
becomes unaffected by the spiral  perturbation.  The two functions are given
in Appendix B of \citet{lin69}.

\noindent The phase of the spiral pattern $\chi$ is defined by
\begin{equation}
\chi=\omega t - m \theta + \Phi(R)
\end{equation}

\noindent  which, in  the case  of  logarithmic spirals  with $\Phi(R)=  m\,
\mathrm{cotg}\, i\, \ln R$ in Eq.~\ref{e:perturbation} ($i$ being the pitch
angle), can be written in a more convenient form
\begin{equation}
\chi = \chi_0 + m (\mathrm{cotg}\, i\, \ln(R/R_0)-(\theta-\theta_0))\,.
\label{e:def2}
\end{equation}

\noindent The subscripts $0$ in the previous equation refer to the value
at the Sun's location. The radial wave number $k$ is then given by
\begin{equation}
k(R)=\Phi'(R)= m\, \mathrm{cotg\,} i / R \,,
\label{e:def3}
\end{equation}
with $k(R)<0$ for trailing waves and $k(R)>0$ for leading waves.

The mean  velocities of  Eq.~\ref{e:basis} vary in  the Galactic plane  as a
function of $R$ and $\theta$, the modulation depending on the mass model via
$\Omega$  and $\kappa$,  the  velocity dispersion  in  the radial  direction
$\sigma_R$ via  the reduction  factors and on  the parameters of  the spiral
perturbation. These parameters are the  number of arms $m$, the amplitude of
the  perturbation $A$,  its pattern  speed $\Omega_{\mathrm  p}$,  the pitch
angle $i$ and  the phase $\chi_0$. In addition we  chose to include $\mvrr0$
as  a free  parameter  while computing  the  model to  account for  possible
uncertainties on $U_{\mathrm{LSR}}$.  While  this parameter is usually taken
into  account  while  computing   the  velocities  ($V_R,V_\theta$)  of  the
observations, here we  include it as a correction to  the predicted \mvr and
$\langle V_\theta  \rangle\,$ in the model  to avoid the  computation of the
velocity field  at each step which  is time consumming. In  the remainder of
the   paper,  we   will  denote   $\P$  the   vector  of   model  parameters
$\P=(m,A,\Omega_{\mathrm p},i, \chi_0,\sigma_R, \mvrr0)$.

For the  rotation curve  of the Milky  Way, we  use the models  I and  II of
\citet{BT08}  Table~2.3, based  on the  mass models  of \citet{dehnen1998b}.
These  models reproduce  equally  well the  circular-speed  curve and  other
observationally  constrained  quantities such  as  the  Oort constants,  the
surface mass density within 1.1 kpc  or the total mass within 100~kpc of the
Milky Way.  The two models correspond to two limiting cases where either the
disc or the halo dominates the rotation curve (model I and II respectively).
The models being computed for $R_0=8$~kpc and $V_{co}\sim 220$~\tkms, we use
the same values  for these two parameters when  computing the galactocentric
velocities in Fig.~\ref{f:fields}.

In  practice,  our  tests  showed   that  all  our  solutions  converged  on
approximately the same value for $\sigma_R$. This is due to our sample being
dominated by the old thin disc  population and we chose to fix $\sigma_R$ to
the best  fit value,  $\sigma_R=34.2 \kms$, to  reduce the dimension  of our
parameter space.  This value of the  dispersion fits very  well the observed
dispersion from the  RAVE sample, excluding the tails  representative of the
thick disc  population and of large  proper motion errors.  Hence, the model
parameters   we    consider   in   the    remainder   of   the    paper   is
$\P=(m,A,\Omega_{\mathrm p},i, \chi_0, \mvrr0)$.

The mean velocities of Eq.~\ref{e:basis} are compared to the two dimensional
velocity  field of  Section~\ref{s:data}.  The  comparison is  done  using a
chi--square estimator
\begin{equation}
\chi^2=\sum_i \frac{(\,\langle V_R \rangle_{i,obs}\,-\langle V_R \rangle_{i,model}\,)^2}{\sigma^2_{i,obs}}\, ,  
\end{equation}

where  the   sum  is  on   all  bins  containing   at  least  5   stars  and
$\sigma_{i,obs}$   is   the   error   on   the  mean   velocity   shown   in
Fig.~\ref{f:fields}  right panel.   We  restrict the  analysis  to the  mean
velocity in  the radial direction, the tangential  velocities being affected
by the asymmetric drift which can  not be properly taken into account in the
model, our sample being a  mixture of stellar populations of different ages,
even though it is dominated by the old thin disc (see Section~\ref{s:data}).

We  note also  that  systematic  distance errors  would  affect the  results
presented  below.   As  shown   in  the  first  paper  \citep{siebert11},  a
systematic error in  the distances affects the measured  gradient in \mvr by
approximately  the same  factor:  a 20\%  overestimate/underestimate of  the
distances induces a $\sim20$\% overestimate/understimate of the amplitude of
the  velocity  gradient  in  \mvr,  which  to  first  order,  results  in  a
higher/lower amplitude  $A$ of the  spiral perturbation by the  same amount.
However, as shown in the same paper, an independent estimate of the velocity
field using red clump stars, for  which an unbiased distance estimate can be
obtained,  shows  a  good  agreement  of  the  velocity  fields,  giving  us
confidence that  our distances  can not be  strongly affected by  an unknown
bias and  we will  not consider  this possibility in  the remainder  in this
paper.

%
%
\section{Results and discussion}
\label{s:result}

\subsection{Parameter space sampling}
\label{s:paramspace}

The  number of arms  in the  Milky Way  is not  known with  certainty.  Both
2-armed and 4-armed spiral pattern are considered in the literature although
an $m=2$ mode  in the stars seems  to be favoured.  In the  analysis we will
consider  both cases  and look  for the  best matching  solution  for either
number of spiral arms.

Also, some  recent works have suggested  that the local standard  of rest is
not  on a  circular  orbit,  eg.  $U_{\mathrm{LSR}}$  might  not be  0~\tkms
\citep{smith2009,smith2012,monibidin2012}.    Therefore,   we  include   the
possibility for a non-null $U$ component of the LSR in the model.

For   the    minimisation,   we    considered   the   standard    value   of
$U_{\mathrm{LSR}}=0\kms$\footnote{Recall   that   $U_{\mathrm{LSR}}   \equiv
  -\mvrr0$.} as well as values of $\pm5$~\tkms whose amplitude correspond to
the  finding of  \citet{smith2012}.   Finally  we left  $\mvrr0$  as a  free
parameter in  the fit. However, we  note that our sample  reaches only 2~kpc
away from the Sun, not deep enough in the plane to disentangle the effect of
a radial  motion of the LSR from  uncertainties in the $U$  component of the
solar motion with respect to the LSR.  Therefore we should keep in mind that
a non-null  best fit  value of the  $\mvrr0$ parameter does  not necessarily
imply a radial motion of the LSR.

\begin{figure*}
\centering
\includegraphics[width=8cm]{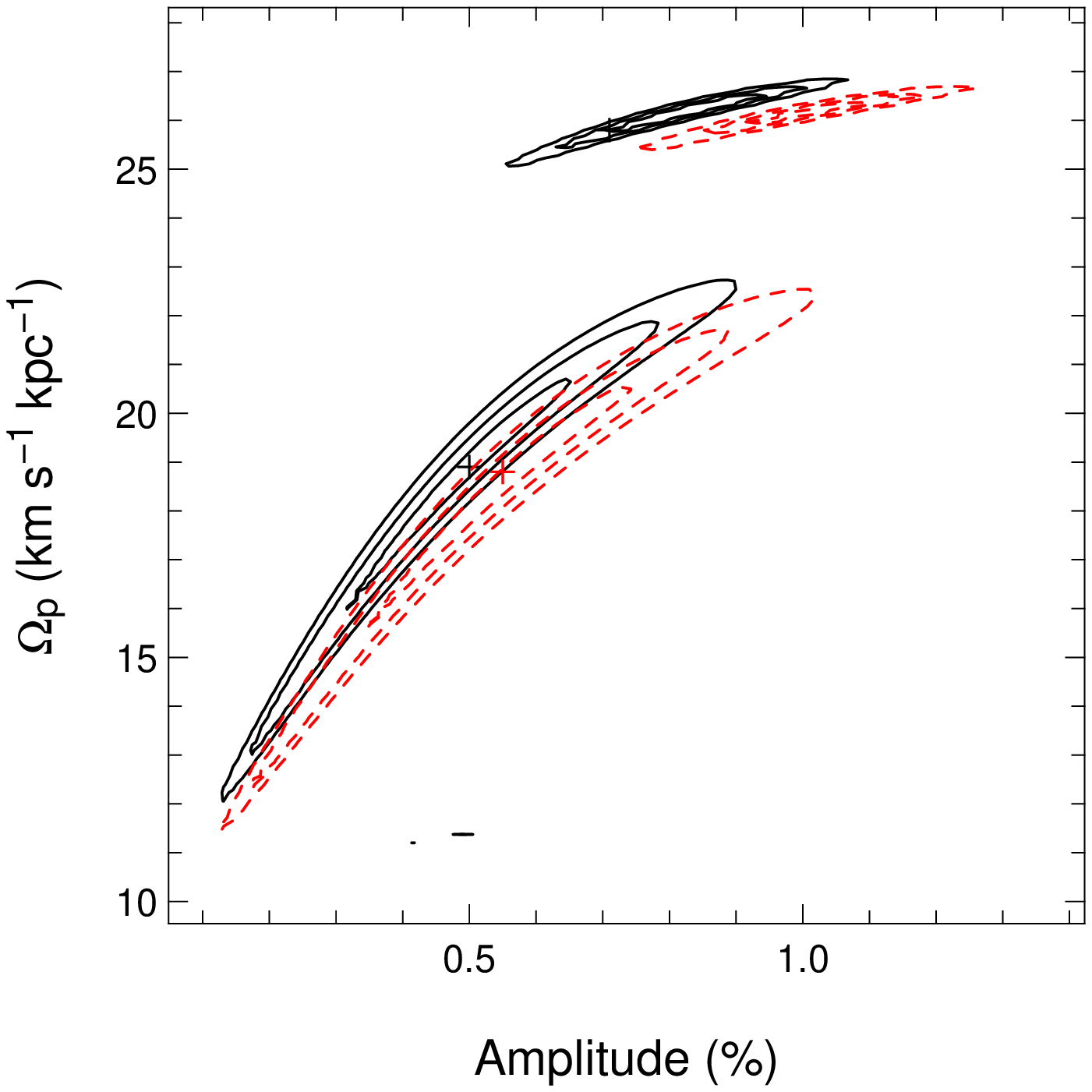}
\includegraphics[width=8cm]{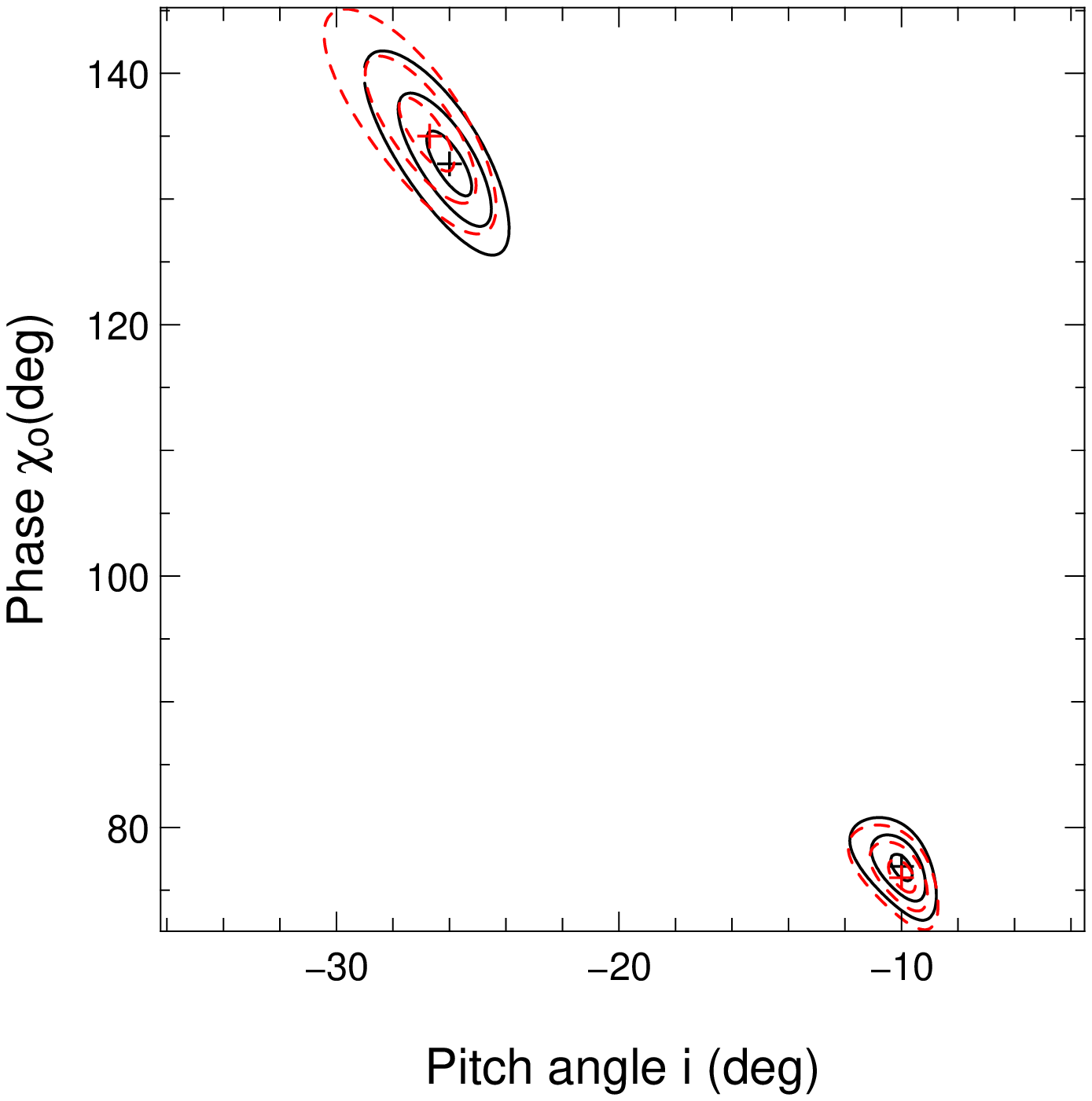}
\caption{Cuts through the chi-square space around the best fit models in the
  amplitude vs  pattern speed  plane (left panel)  and phase vs  pitch angle
  plane (right panel).  The crosses mark  the location of the best fit while
  the contours are the 1, 2 and 3 $\sigma$ limits. The amplitude in the left
  panel is  given as a percentage  of the background potential  at the Sun's
  location.  The plain  lines are for the mass model I  and the dashed lines
  for the mass model II.  In both panels the top contours are for $m=4$, the
  bottom contours for $m=2$.}
\label{f:likelihood}
\end{figure*}

The summary  of the chi-square analysis is  presented in Table~\ref{t:model}
and  the  chi-square   contours  for  the  best  models   are  presented  in
Fig.~\ref{f:likelihood}. In  this figure, the two  panels show the  1, 2 and
3-$\sigma$  contours,  fixing all  the  other  parameters  to the  best  fit
solution, in  the $\Omega_p$ versus  amplitude plane (left panel)  and pitch
angle $i$ versus phase $\chi_0$ (right  panel).  The plain lines are for the
mass model I, the red dashed lines for model II.

\begin{table*}
  \caption{Chi-square results. Parameters with error bars were kept
    free in the minimisation. The error bars correspond to the 1-$\sigma$
    internal errors obtained from the chi-square contour. Models marked with
    (*) have a large pitch angle (open arms) and do not satisfy the
    tight-winding approximation. The number of pixels used in the
    minimisation procedure is 1595.}
\label{t:model}
\begin{tabular}{c c c c c c c c c l}
\hline
Mass  & $m$ & $\mvrr0$    & $\Omega_p$         & $A$ & $i$ &  $\chi_0$ &
$\chi^2$&\\
model &     & \tkms       & $\kmskpc$         & \% (total, disc) & deg & deg &&\\
\hline
I & 2 & $0.9^{+0.1}_{-0.1}$ & $18.9^{+0.3}_{-0.2}$ & $(0.50^{+0.02}_{-0.02},2.27^{+0.08}_{-0.07})$ & $-10.0^{+0.4}_{-0.4}$ & $76.9^{+1.1}_{-1.2}$ & 1829.00&\\
I & 2 & -5                 & $16.1^{+0.1}_{-0.1}$ & $(0.78^{+0.01}_{-0.01},3.50^{+0.07}_{-0.06})$ & $-23.2^{+0.3}_{-0.5}$ & $57.3^{+0.6}_{-0.5}$ & 1943.14&(*)\\
I & 2 & 0                 & $18.8^{+0.2}_{-0.3}$ & $(0.49^{+0.02}_{-0.02},2.21^{+0.08}_{-0.09})$ & $-9.1^{+0.3}_{-0.4}$ & $65.8^{+1.5}_{-1.0}$ & 1831.99&\\
I & 2 & 5                & $19.3^{+0.1}_{-0.2}$ & $(0.69^{+0.02}_{-0.01},3.12^{+0.09}_{-0.05})$ & $-15.6^{+0.7}_{-0.6}$ & $112.1^{+1.0}_{-0.9}$ & 1853.04&\\
II& 2 & $0.9^{+0.3}_{-0.2}$ & $18.6^{+0.3}_{-0.2}$ & $(0.55^{+0.02}_{-0.02},3.09^{+0.10}_{-0.13})$ & $-10.0^{+0.4}_{-0.4}$ & $76.0^{+1.3}_{-1.0}$ & 1828.46&\\
II& 2 & -5                 & $15.1^{+0.1}_{-0.1}$ & $(0.71^{+0.01}_{-0.01},3.93^{+0.06}_{-0.07})$ & $-22.3^{+0.3}_{-0.6}$ & $55.7^{+0.7}_{-0.4}$ & 1940.75&(*)\\
II& 2 & 0                 & $18.5^{+0.3}_{-0.2}$ & $(0.54^{+0.02}_{-0.03},3.03^{+0.09}_{-0.15})$ & $-9.3^{+0.4}_{-0.3}$ & $66.6^{+1.3}_{-1.1}$ & 1830.17&\\
II& 2 & 5                & $16.2^{+0.1}_{-0.1}$ & $(0.51^{+0.01}_{-0.01},2.84^{+0.07}_{-0.05})$ & $-16.3^{+0.7}_{-0.6}$ & $111.3^{+0.9}_{-0.8}$ & 1859.98&\\
I & 4 & $4.9^{+0.1}_{-0.1}$ &$25.8^{+0.1}_{-0.1}$ & $(0.71^{+0.02}_{-0.01},3.22^{+0.08}_{-0.06})$ & $-26.0^{+0.6}_{-0.5}$ & $132.8^{+1.8}_{-1.7}$ & 1833.52&(*)\\
II& 4 & $4.9^{+0.1}_{-0.1}$ & $25.9^{+0.1}_{-0.1}$ & $(0.91^{+0.03}_{-0.02},5.09^{+0.16}_{-0.09})$ & $-26.7^{+1.0}_{-0.4}$ & $135.0^{+2.3}_{-1.7}$ & 1829.55&(*)\\
\hline
\end{tabular}
\end{table*}

The best  fit model is obtained  for a two  armed spiral mode with  the mass
model  II.  The  best  fit  solution for  model  I is  equally  good with  a
chi-square  difference of 0.5.   Generally, the  difference between  the two
mass models  is low.  This is expected  as within the region  sampled by the
RAVE       data,       the       models      are       comparable       with
$\Omega_{\mathrm{mI}}-\Omega_{\mathrm{mII}}   \approx   0.35\kmskpc$.    The
density wave model  being not sensitive to the details of  the mass model --
the latter  entering the equations  only through $\Omega(R)$  and $\kappa$--
the  two  models  can  only  be  distinguished in  regions  where  they  are
significantly different (for example closer to the Galactic centre).

The chi-square  value of the best $m=4$  solution is also close  to the best
$m=2$  solution.  However,  for $m=4$  the pitch  angle $i$  is found  to be
$\approx-26$ degrees,  out of the bounds of  the tight-winding approximation
upon which  the density wave model relies  ($max(|i|)\approx 15-20$ degrees,
\citet{lin69}).   Hence,  for a  four-armed  pattern,  we  conclude that  no
satisfactory solution  is found and we  will discard four  armed patterns in
the following discussion.

Focusing on  the $m=2$  mode, a strong  correlation is observed  between the
amplitude and the pattern speed  while the correlation is weaker between the
pitch angle and the phase (Fig.~\ref{f:likelihood}).  If the phase and pitch
angle are  well determined,  the shape of  the contours in  amplitude versus
$\Omega_p$ indicate  a large range  in possible solutions at  the 3-$\sigma$
level: $\Omega_p$ varies  from 12 to 22~\tkmskpc and  the amplitude from 0.1
to 0.9\% of the background potential. The amplitude of the best fit model is
$A=0.55\%$ of the background potential. This value translates to 14\% of the
background density which is close  to the value proposed by \citet{MF10} and
consistent with earlier  determinations summarized in \citet{antoja2011} for
the   local  spiral  amplitude.    We  also   note  that   corotating  waves
($\Omega_p\sim27.5\kmskpc$) seem to be excluded.

Finally,  among the  $m=2$  solutions, a  low  radial component  of the  LSR
velocity  is prefered.   The best  fit model  converges  to $\mvrr0=0.9\kms$
while  a zero  radial component  can  not be  ruled out  when comparing  the
chi-quare values.  On  the other hand, a more  pronounced outwards motion of
the LSR ($\mvrr0=5\kms$) shows  significantly larger chi-square values while
an inwards motion of $-5\kms$ as suggested by \citet{smith2012} is even less
consistent. However in  the latter case, the model  converges outside of the
range of allowed values for the pitch angle which limits the conclusions one
can draw  from this result.  We note  that our best fit  value is consistent
with  the  \citet{schoenrich2010}  errors   at  the  2-$\sigma$  level  when
considering   the   error  bars   on   the   determination   of  $U$   (e.g.
$U=11.1^{0.69}_{-0.75} \kms$).  In the  next section, we will concentrate on
the  best $m=2$  models  and their  implications  for the  structure of  the
Galactic disc.

\subsection{Resonances and spiral structure}

\begin{figure*}
\centering
\begin{tabular}{c c}
\includegraphics[width=7cm]{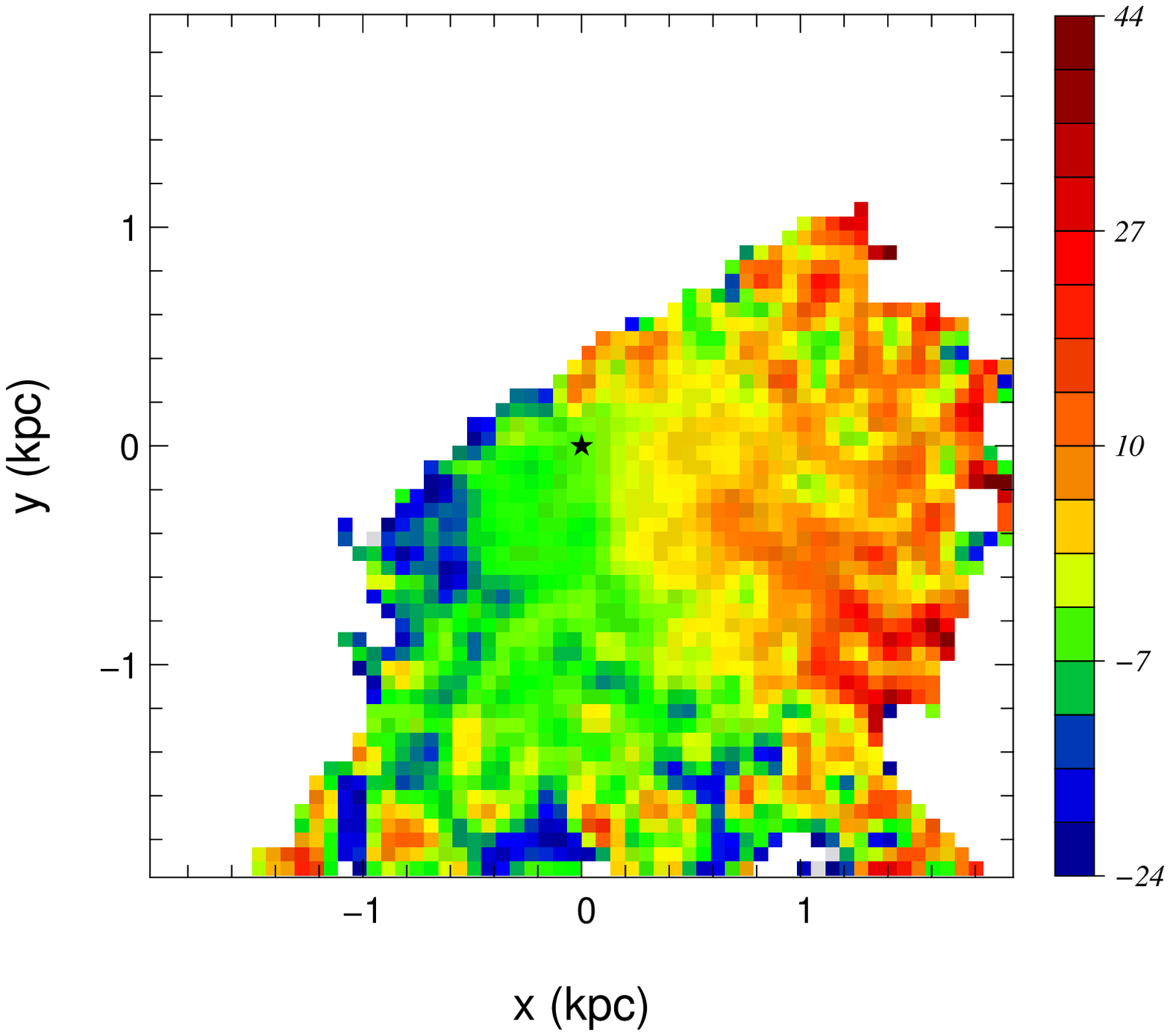} &
\includegraphics[width=7cm]{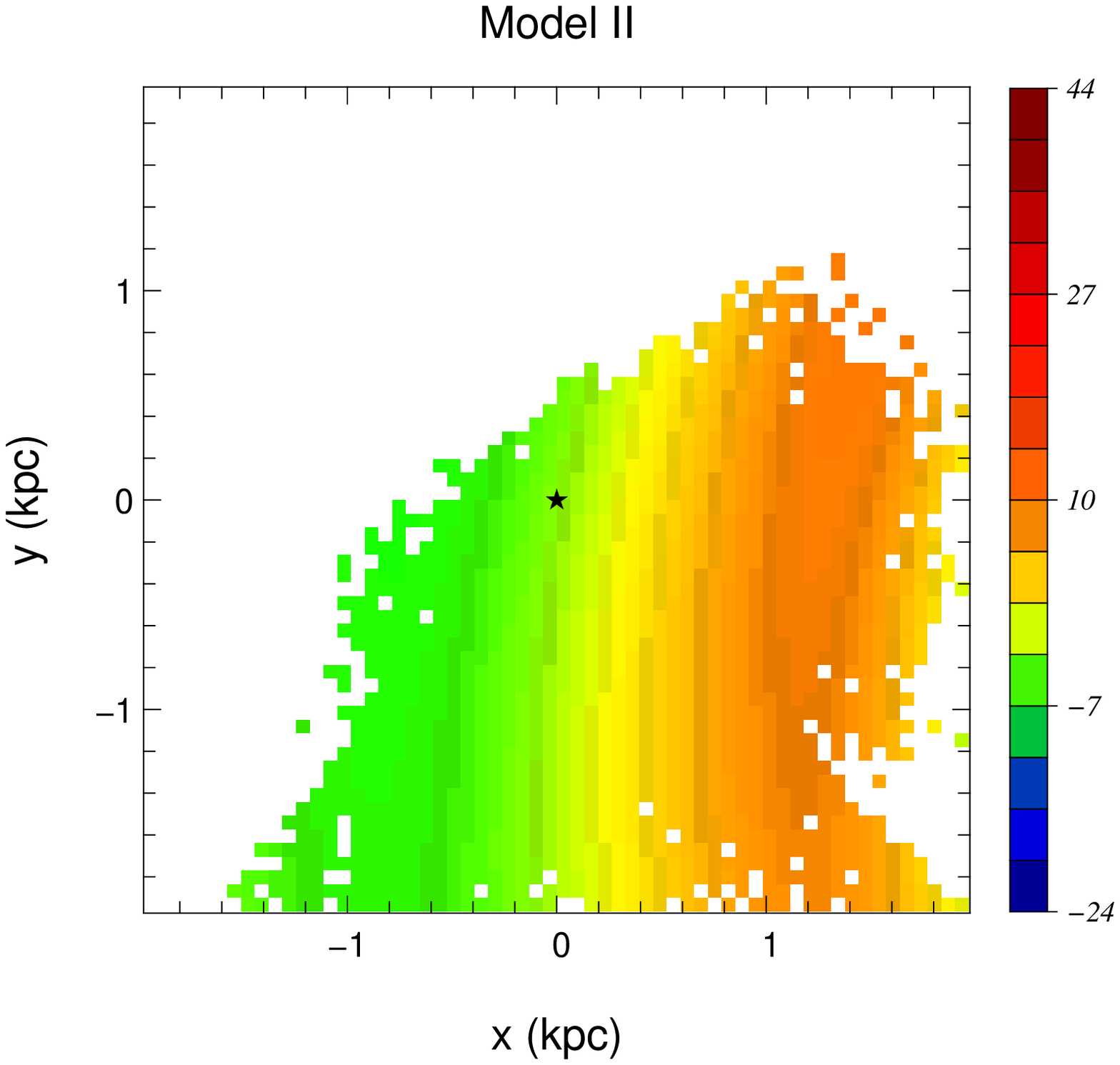}\\
\includegraphics[width=7cm]{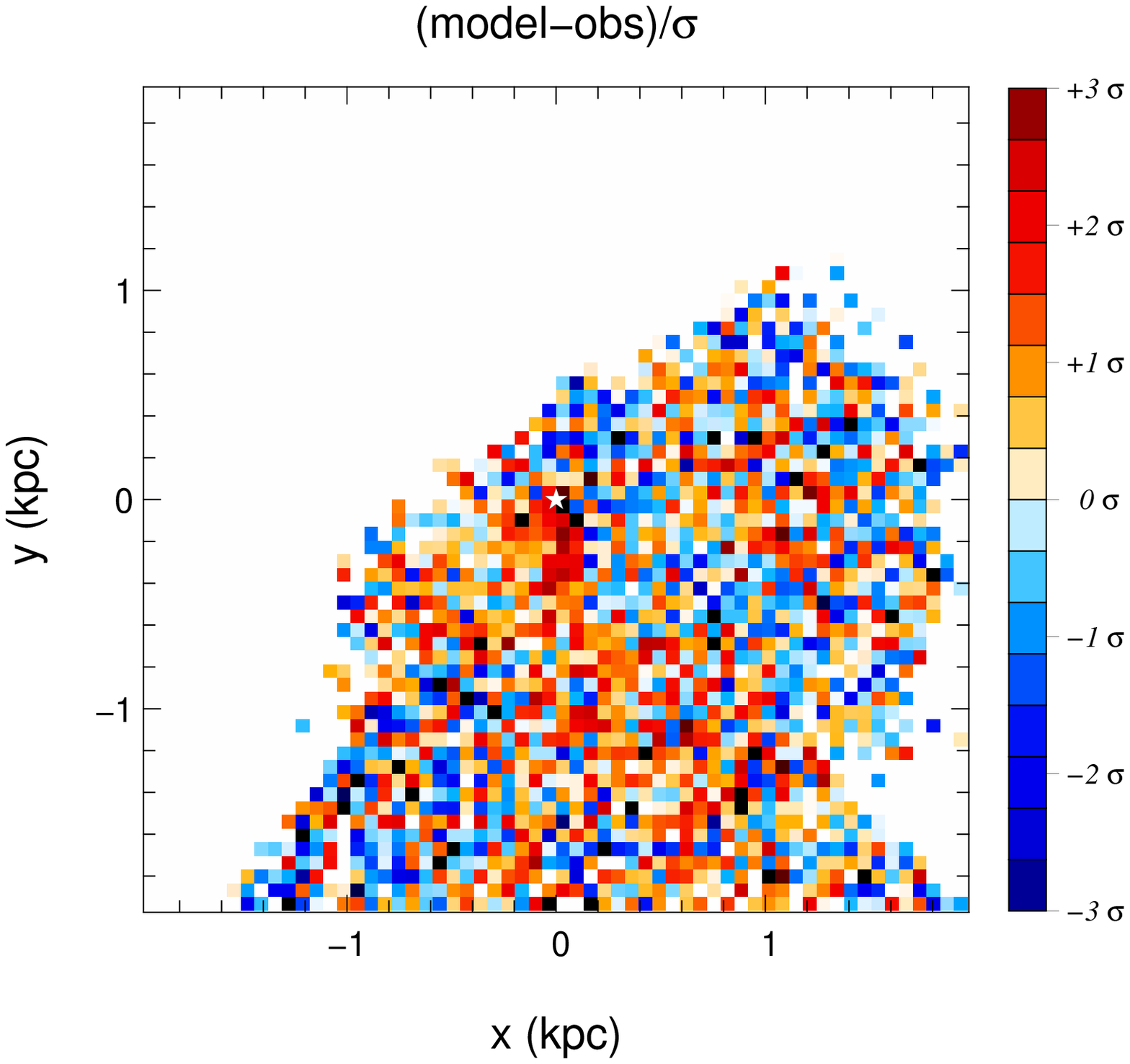}&
\includegraphics[width=7cm]{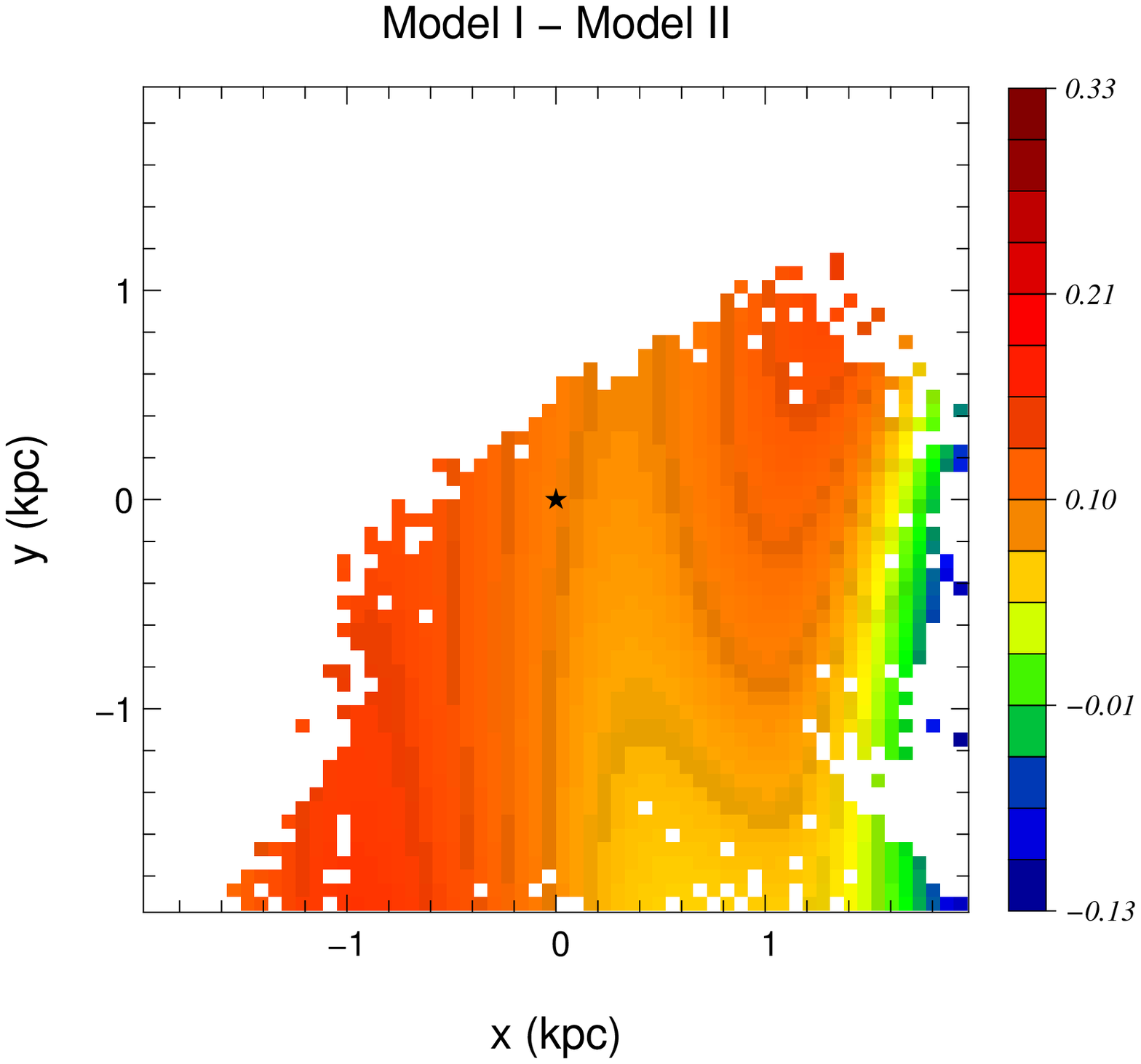}\\
\end{tabular}
\caption{Top panels: observed velocity field (left) and model velocity field
  for the best  fit solution using the mass model  II (right). Bottom right:
  velocity field  difference between the  best fit solutions using  the mass
  model  I and  II.  The  colour  coding follows  the median  galactocentric
  radial velocity  in \tkms in the  three panels.  Bottom  left : difference
  $\langle V_{R,mII}  \rangle-\langle V_{R,observed} \rangle$  normalised by
  the observational errors showing that  all velocities on the 2 dimentional
  map  are well  recovered within  the observational  uncertainties.  On all
  panels, the Sun's location is at (0,0) and is marked by the black or white
  star.}
\label{f:velfield}
\end{figure*}

The best  fit \mvr velocity  fields for the  $m=2$ solutions using  the mass
models II  is presented  in Fig.~\ref{f:velfield}.  Although  both reproduce
equally well  the structure  of the velocity  field between  $y=\pm1$~kpc, a
small difference between the two solutions exists, model I predicting larger
velocities in  the top right  and lower left  corners of our  sample (bottom
right  panel).  However the  velocity difference  reaches only  0.2~\tkms, a
level  that lies  below  the capabilities  of  our data.   The region  below
$y=-1$~kpc and  at $0<x<1$~kpc is  apprently poorly reproduced,  however the
lower left panel of  Fig.~\ref{f:velfield} indicates that our solution stays
well   within    the   observational    errors.    The   right    panel   of
Fig.~\ref{f:fields} indicates  that this region suffers  from large velocity
errors and  has therefore a  lower weight in  the solution.  In  the regions
where  our data are  of the  best quality  (mostly $|y|<1$~kpc),  our models
reproduce adequately the observed velocity field.

\begin{figure*}
\centering
\includegraphics[width=8cm]{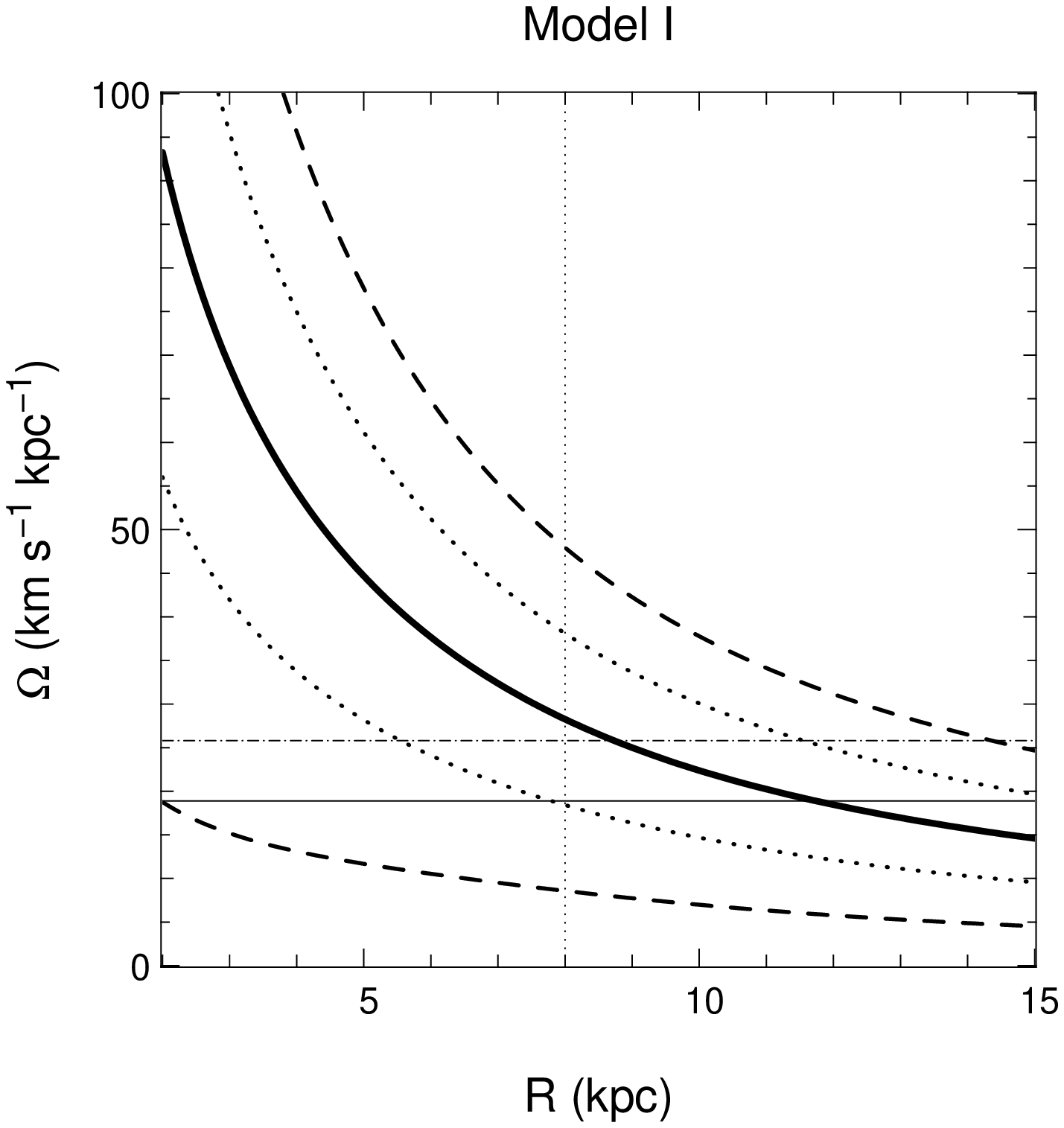}
\includegraphics[width=8cm]{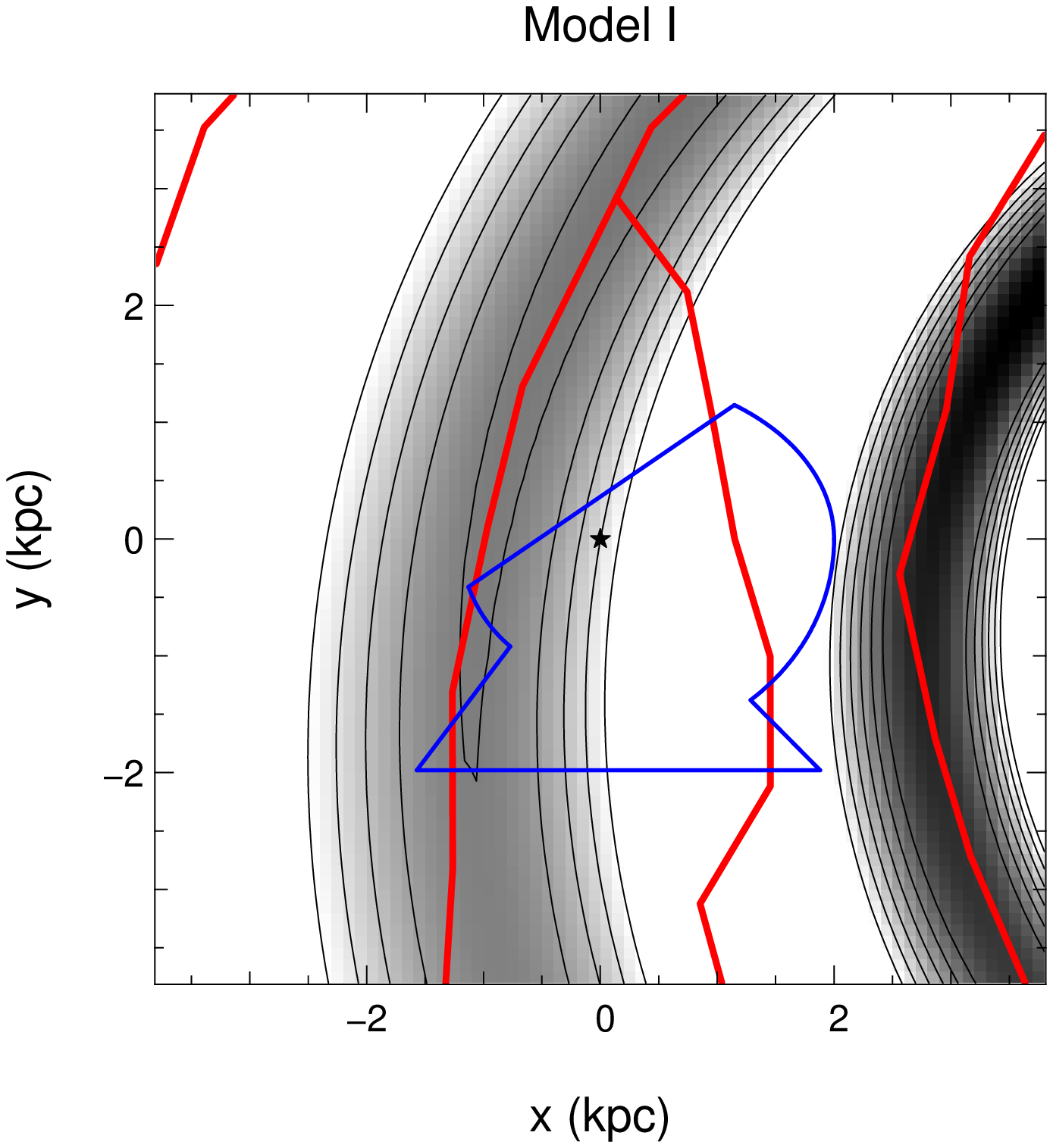}
\caption{Left: circular  frequency as a function  of galactocentric distance
  for the mass  model I of Binney \& Tremaine (thick  lines). The dotted and
  dashed  lines  are  respectively  the  relations  $\Omega\pm\kappa/4$  and
  $\Omega\pm\kappa/2$ versus  R. The  location of the  Sun is marked  by the
  vertical dotted line at $R_0=8\kpc$.  The horizontal lines are the pattern
  speed corresponding  to the  best fit models  for $m=2$ (plain  lines) and
  $m=4$  (dash-dotted  lines).   Right:   Density  associated  to  the  best
  model. The grey shading  and contours represent the overdensity associated
  to  the   spiral  perturbation.   The   contours  are  evenly   spaced  by
  $0.1\Sigma_0$, the  background column density, from  0.1 to 0.5$\Sigma_0$.
  The blue  contour depicts the  footprint of the  RAVE data. The  red lines
  mark the  location of the spiral  arms in the  gas from \citet{englmaier}.
  From left to right we have  Cygnus arm (top left corner), the Perseus arm,
  the Sagittarius-Carina arm and  the Scutum-Centaurus arm.  The results for
  the mass model II are almost identical and are therefore not presented.}
\label{f:resonances}
\end{figure*}

Focusing  on the  pattern speed,  our best  models suggest  that the  Sun is
located  $\sim200\,$pc  inside   the  inner  4:1  resonance  (ultra-harmonic
renonance  or UHR)  of  the spiral  pattern (Fig.~\ref{f:resonances}).   Our
finding for the pattern speed $\Omega_p=$18--19\tkmskpc is in agreement with
recent   studies   that   also   place   the   Sun   close   to   the   UHR:
$\Omega_p=17\kmskpc$   by    \citet{antoja2011},   $\Omega_p=18\kmskpc$   by
\citet{quillen2005} or $\Omega_p/\Omega_0=0.65$ by \citet{pompeia2011} to be
compared to $\Omega_p/\Omega_0\sim0.68$ in  our study.  However, as shown by
\citet{gerhard2011}, determinations of the  spiral arms' pattern speed range
from 17  to 28 \tkmskpc,  the higher values  being prefered by  open cluster
birthplaces while  hydrodynamical simulations and  phase space substructures
favour slower  pattern speeds.  It is  interesting to note  that the pattern
speeds       found       from       velocity       space       substructures
\citep{quillen2005,antoja2011,pompeia2011}  are close  to  our value.   This
would indicate a  similar origin for the velocity  gradient and the velocity
substructures,  reinforcing our  assumption  that the  velocity gradient  we
observed  is due  to spiral  arms.  However  this statement  must be  put in
perspective as  both types of  study rely on  the same assumptions  that the
spiral pattern is long-lived and tightly wound.

Comparing the predicted  density pattern to the location  of the spiral arms
obtained  by  \citet{englmaier}  in  the  gas,  we  find  a  good  agreement
(Fig.~\ref{f:resonances}  right  panel).   Both  the  Perseus  arm  and  the
Centaurus arm are recovered  at the proper location.  The Sagittarius-Carina
arm is not recovered in our  models. This indicates that this feature is not
a dominant feature in the  Solar neighborhood, reinforcing the view that the
Milky Way  spiral arm  pattern is  dominated by two  main arms,  Perseus and
Centaurus        \citep{drimmel2000,drimmel2001,benjamin2005,churchwell2009}.
Contrary to the $m=2$ models, our best fit $m=4$ solution does not reproduce
any of the known spiral arms,  in addition to being invalidated by its large
pitch angle, hence an $m=2$ mode for  the spiral pattern in the Milky Way is
preferred within the Lin-Shu regime.

%
%
\section{Conclusion} 

We have  analysed the velocity gradient detected  by \citet{siebert11} using
the RAVE data in the framework  of the density wave model of \citet{linshu},
assuming that the  velocity gradient we detected is due  only to spiral arms
and that the spiral arms in the Milky Way are long-lived.

Our model converges  properly for an $m=2$ pattern,  while if the chi-square
of  the $m=4$ solutions  are comparable,  the predicted  pitch angle  is too
large, invalidating the solution.

The best fit solutions for  $m=2$ reproduce adequately the observed velocity
field  for \mvr  in  the region  $|y|<1$~kpc  where our  data  are the  most
reliable. Outside of  this region, the difference between  the model and the
observations is still within the observational errors although the agreement
is less clear.

The predicted  pattern speed places the  Sun about 200~pc  outside the inner
UHR  of the spiral  arms.  Such  a location  of the  Sun is  consistent with
previous works  based on velocity space substructures,  suggesting a similar
origin for the velocity space substructures and the \mvr gradient.  Our best
fit    value   for    the    amplitude   of    the   spiral    perturbation,
$A=0.55^{+0.02}_{-0.02}\%$  of  the  background  potential or  14\%  of  the
background  density,  is  consistent  with  the  value  proposed  by,  e.g.,
\citet{MF10} and is also in  the range of earlier measurements as summarized
in \citet{antoja2011}.

Comparing our  model to the location  of spiral arms  in the gas, we  find a
good  agreement  with  the  location  of  the major  spiral  arms  given  by
\citet{englmaier}.   The density  enhancement  predicted by  our best  model
matches the location of the  Perseus and Centaurus arms. The Sagittarius arm
is not reproduced by our  solution which tends to reinforce previous studies
concluding that the  Milky Way spiral potential is  dominated by a two-armed
mode, the Sagittarius-Carina  arm being a minor feature  for the dynamics of
the disc.

Our study relies on the density  wave model of \citet{linshu} and we assumed
no vertical variation of the \mvr  field within the limit of our data.  RAVE
data do contain  the 3-dimensional spatial information which  we will use in
further studies.   However, going from 2D  to 3D requires an  upgrade of our
modeling technique taking properly into account the asymmetric drift and the
vertical   variation   of   the   spiral  potential.    Moreover,   vertical
perturbations leading to possible  variations of $\langle V_z \rangle (R,z)$
\citep{smith2012,widrow,williams} are  intrinsically not taken  into account
in  our analysis.   Future 3D  simulations of  such perturbations  and their
possible influence on the \mvr  field will be necessary to disentangle their
possible effects  from the velocity  gradient modeled here. Finally  we note
that  our model  is local  as a  result of  the  tight-winding approximation
\citep[see  for example  discussion in][Section  1.4.2]{binney2012}. Ongoing
surveys like Gaia-ESO (GES) or  SDSS/SEGUE will provide data in the Galactic
plane that  can be used to test  our models further in  towards the Galactic
centre (GES)  or further  out (SDSS/SEGUE). It  will be interesting  to test
whether  these two surveys  predict the  same pattern  speed for  the spiral
arms.

\section*{Acknowledgements}
Funding  for  RAVE  has   been  provided  by:  the  Australian  Astronomical
Observatory;  the  Leibniz-Institut  fuer  Astrophysik  Potsdam  (AIP);  the
Australian National University; the  Australian Research Council; the French
National Research Agency;  the German Research Foundation (SPP  1177 and SFB
881); the European Research Council (ERC-StG 240271 Galactica); the Istituto
Nazionale  di  Astrofisica at  Padova;  The  Johns  Hopkins University;  the
National  Science Foundation  of the  USA  (AST-0908326); the  W.  M.   Keck
foundation; the  Macquarie University;  the Netherlands Research  School for
Astronomy; the Natural Sciences  and Engineering Research Council of Canada;
the Slovenian  Research Agency; the  Swiss National Science  Foundation; the
Science  \& Technology  Facilities Council  of the  UK;  Opticon; Strasbourg
Observatory; and  the Universities of Groningen, Heidelberg  and Sydney. The
RAVE web site is at http://www.rave-survey.org.



%
\bibliographystyle{aa}


\end{document}